\input amstex
\documentstyle{amsppt}
\mag=\magstephalf
\pageno=1

\TagsOnRight
\hcorrection{0.5truecm}
\interlinepenalty=1000
\hsize=6.5truein
\voffset=10pt
\vsize =9.8truein
\advance\vsize by -\voffset
\nologo

\NoBlackBoxes
\define \ii{{\roman i}}
\define \ee{{\roman e}}

\define \dd{{\roman d}}

%

\topmatter
\title
\endtitle
{\centerline{\bf{Quantum Coupled Nonlinear Schr\"odinger 
System with Different Masses}}}
\author
\endauthor
\affil
 Shigeki MATSUTANI\\
 2-4-11 Sairengi, Niihama, Ehime, 792 Japan \\
\endaffil
\topmatter
\title
\endtitle
{\centerline{\bf{}}}

\abstract 

In this letter, I have considered one-dimensional quantum
system with different masses  $m$ and $M$, which does not 
appear integrable in general. However  I have found  an exact
 two-body wave function and
due to the extension of the integrable system to more 
general system, it was concluded that the rapidity or 
quasi-momentum in the 
integrable system should be regarded as a modification
 of velocity rather than that of momentum.
I have also considered the three-body wave function 
and argued its integrable condition.

\endabstract
\endtopmatter



Recently  solitary waves in a classical particle chain
 consisting of different masses were discovered [1,2].
In the  one-dimensional classical particle chain with
 alternate different masses $m$ and $M$ whose ratio is
  a certain value given by the cyclotomic polynomial, 
there is a solitary wave for an appropriate initial condition.

In this letter, I will consider a kind of quantum version of 
 such system.

\vskip 0.5 cm

Here, I will investigate the non-relativistic one-dimensional
 quantum system,
$$
      \left(-\sum_{i=1}^N\partial_{x_i}^2 
      -\gamma\sum_{j=1}^n  \partial_{y_j}^2+
       c\sum_{j=1}^n\sum_{i=1}^N\delta(x_i-y_j)\right)
       \Psi(x,y)  =E\Psi(x,y), \tag 1
$$
where $1/\gamma$ is a normalized mass, $\gamma\ge1$.
By an appropriate scale transformation, this system can be
 interpreted as quantizing particles of masses $m$ and $M$
  with   the relation $M/m=\gamma$.
This system is a generalization of the integrable hard
 core boson system with the same mass ($\gamma=1$) [3-6].
Even though the system given by (1) does not appear integrable
 for  general mass ratio $\gamma$,
 I expect that  by investigating such exotic system, physical
  meaning of the integrable system might be revealed.

In this letter, I will study the system (1) parallelling to
 the argument of Thacker for the 
integrable system ($\gamma=1$) in [5-6].
For later convenience, I will express this equation (1) by
 the second quantized field,
$$
  H =\int \dd x \left(\partial_x \Phi^*\partial_x \Phi+
   \gamma \partial_x \phi^* \partial_x \phi +  
          c\phi^*\phi \Phi^*  \Phi \right), \tag 2
$$
where $\phi$ and $\Phi$ are non-relativistic bosonic fields
with canonical equal time commutation relations
$$
               [\Phi(x,t),\Phi^*(y,t)]=\delta(x-y),\quad
               [\phi(x,t),\phi^*(y,t)]=\delta(x-y) ,
               \tag 3
$$
and the other relations which are commutative. 
For $\gamma=1$ case, this is essentially the same as
 the nonlinear Schr\"odinger system [3-6].

Hamiltonian commutes with the particle number operators
$\hat  N := \int \Phi^* \Phi \dd x$ and
 $\hat n:=\int \phi^*\phi \dd x$.
Thus I can precisely deal with $n$ and $N$ particles state,
 where $N$ and $n$ are the eigenvalues of the operators
  $\hat N$ and $\hat n$ respectively.
Let $k$ be a momentum eigenstate of $\Phi$ particle and
 $k/\sqrt{\gamma}$ be
that of $\phi$. I will introduce the creation operators,
$$
    A^\dagger_k= \ \int \dd x \ee^{\ii k x} \Phi^*(x),
      \tag 4.a
$$
and
$$
    a^\dagger_l= \ \int \dd y \ee^{\ii l y} \phi^*(y).
     \tag 4.b
$$
The non-perturbative momentum state will be denoted as
$$
      |k_1,\cdots,k_N,k_{N+1}/\sqrt{\gamma},
      \cdots,k_{n+N}/\sqrt{\gamma}>:=
     \prod_{i=1}^N A^\dagger_{k_i} \prod_{j=N+1}^{n+N}
      a^\dagger_{k_j/\sqrt{\gamma}}|0>, \tag 5
$$
where $|0>$ is the vacuum state of this system.
The total energy $\omega_{k}$ and the total momentum $P_k$
 of the state are expressed as respectively,
$$
     \omega_{k}= \sum_{i=1}^N k_i^2+\sum_{j=1}^n k_j^2 ,
      \tag 6.a
$$
and
$$
     P_k=\sum_{i=1}^N k_i+\sum_{j=1}^n k_j/\sqrt{\gamma} .
      \tag 6.b
$$

Following Thacker's argument of the ordinary quantum nonlinear
 Schr\"odinger equation [6,7],
I will compute the proper states of this system using 
the perturbation theory.
By means of an old fashion perturbative method, an exact
 state is expressed by [6,7]
$$
    \split
   |\Psi(k_1,&\cdots,k_N,k_{N+1}/\sqrt{\gamma},
   \cdots,k_{n+N}/\sqrt{\gamma})>=\\
   &\sum_{l=0}^\infty [G_0(\omega_k)V]^l 
   |k_1,\cdots,k_N,k_{N+1}/\sqrt{\gamma},
   \cdots,k_{n+N}/\sqrt{\gamma}>, \\
               \endsplit \tag 7
$$
where $G_0$ is the Green function operator for free particles,
$$
     G_0(\omega)=\frac{1}{\omega-H_0+\ii \epsilon} \tag 8
$$
and $H_0$ and $V$ are the free terms  and last term in (2)
 respectively.

It is obvious that any states consisting of only $\phi$ or
 $\Phi$ are reduced to free ones. Thus I will be concerned
  only with a mixing state of $\phi$ and $\Phi$. 

\subheading{Two-body collision}

I will consider a non-trivial two-body wave function of
 $\phi$ and $\Phi$ here.

Before I start the quantum computation,
I will mention elastic collision of  two classical particles
 with mass $\gamma$ and 1 
for the initial momentums $k_1$ and $k_2/\sqrt{\gamma}$.
Due to the energy-momentum conservation law, the variation of 
these momentums  at a collision,
 $(k_1,k_2/\sqrt{\gamma}) \to (K_1,K_2/\sqrt{\gamma})$,
 are given as [1,2]
$$
       \pmatrix K_1\\ K_2 \endpmatrix = 
      \Gamma_+
     \pmatrix k_1\\ k_2\endpmatrix , \qquad
      \Gamma_+:=\pmatrix \beta &\sqrt{1-\beta^2} \\ 
      \sqrt{1-\beta^2} &-\beta  \endpmatrix  ,\tag 9
$$
where $\beta := (\gamma-1)/(\gamma+1)$ and then 
$ \gamma \equiv (1+\beta)/(1-\beta)$.

I will study the quantum analogue of above situation (9)
 given through the Hamiltonian (2).
The old fashion perturbative calculation is performed by
 the diagram Fig.1 [5,6]. The first term which corresponds
to the non-perturbative state is 
given by,
$$
  \int \dd x \dd y \ee^{\ii (k_1 x+k_2y/\sqrt{\gamma} )} 
    \phi^*(x) \Phi^*(y) |0>. \tag 10
$$
The second term in Fig.1 becomes
$$
      \split
    \int \frac{\dd p_1}{2 \pi}& \frac{\dd p_2}
    {2 \pi\sqrt{\gamma}}
  \left(\frac{1}{k_1^2+k_2^2-p_1^2-p_2^2+\ii \epsilon}\right)
  2 \pi c\delta(k_1+k_2/\sqrt{\gamma}-p_1-p_2/\sqrt{\gamma}) 
  A^\dagger_{p_1} a^\dagger_{p_2/\sqrt{\gamma}} |0>\\
  &=\frac{c}{k_1-\sqrt{\gamma}k_2}\int \dd x \dd y\\ 
  &\qquad \left[ \theta(y-x)
    \ee^{\ii (k_1 x+k_2 y/\sqrt{\gamma})}
   +\theta(x-y) \ee^{\ii (K_1 x+K_2y/\sqrt{\gamma} )}\right] 
   \Phi^*(x)\phi^*(y) |0>\\ \endsplit , \tag 11
$$
where $k_1>k_2\sqrt{\gamma}$. The higher order terms
 are given as the loop diagrams illustrated in Fig.1 
 and have the same structure, which are calculated as
$$
  c\int\frac{\dd p_1}{2 \pi}\frac{\dd p_2}{2 \pi\sqrt{\gamma}}
  \left(\frac{\delta(k_1+k_2/
  \sqrt{\gamma}-p_1-p_2/\sqrt{\gamma})}
   {k_1^2+k_2^2-p_1^2-p_2^2+\ii \epsilon}\right)
   = \frac{\ii c}{2(\sqrt{\gamma} k_2-k_1)} . \tag 12
$$
Here the series of the diagrams in Fig.1 converges and the
 converged point is defined by $\Delta(k_1,k_2)$,
$$
   1 + 2\left(\frac{\ii c}{2(\sqrt{\gamma} k_2-k_1)} \right)+
    2\left(\frac{\ii c}{2(\sqrt{\gamma} k_2-k_1)} \right)^2+
    \cdots
    =\frac{k_1-\sqrt{\gamma} k_2-ic/2}
    {k_1-\sqrt{\gamma} k_2+ic/2}
     =:\ee^{-\ii \Delta(k1,\sqrt{\gamma}k2)/2}, \tag 13
$$
where $\Delta$ function has the property,
$$
     \Delta(K1,K2)=-\Delta(k1,k2) . \tag 14
$$

Accordingly I obtain the full exact two-body wave function for
 $k_1 <k_2\sqrt{\gamma}$ 
and then $K_2\sqrt{\gamma}>K_1$,
$$
\split
   |\Psi(k_1,k_{2}/\sqrt{\gamma})>=&\int \dd x \dd y
   \left[\theta(x-y)\left(\ee^{\ii(k_1 x+k_2y/\sqrt{\gamma})}+
  (\ee^{ \ii \Delta(K1,K2)/2}-1)\ee^{\ii(K_1 x+K_2y/
  \sqrt{\gamma})}\right)\right.\\
  &+   \left. 
  \theta(y-x)\ee^{\ii(k_1 x+k_2y/\sqrt{\gamma}
  - \Delta(k1,k2)/2)}\right]\Phi^*(x)\phi^*(y)|0>\\
      \endsplit . \tag 15
$$
The constituent parts of this wave function are illustrated 
as Fig.2.
In the region of $x<y$, there are incident state and 
reflection state while the state in  the region $y<x$ is
 the transparent one.
If $\gamma=1$ and $\phi=\Phi$, this wave function becomes 
the ordinary two-body state of quantum non-linear Schr\"odinger 
system [3-6].

It should be noted that the phase factor of the wave function, 
which  constitutes the S-matrix, is determined by the velocity 
$k_1$ and $\sqrt{\gamma} k_2$ rather than the momentum $k_1$
 and $k_2/\sqrt{\gamma}$. 
Hence though the spectral parameters $k_1$ and $k_2$
which appear in $\Delta(k_1,k_2)$ are sometimes called as
 "quasi-momentum" [3]
for $\gamma=1$ case, they should be interpreted as 
modification of the velocities.

\subheading{Three-body wave function }

Next I will comment upon the computations of  
the three-body wave function  in which
two $\Phi$ particles  have the momentum $(k_1,k_2)$ and a
 $\phi$ particle  has $k_3/\sqrt{\gamma}$. 
Due to the classical properties of the one-dimensional chain 
[1,2], it will be 
clarified that wave function of these three particles can be 
 calculated for a certain mass ratio even though this system 
 is not integrable in general. 

The classical elastic scattering of these particles  are 
also expressed by this transformation matrices like (9).
The collision of 
$(p_1,p_3/\sqrt{\gamma}) \to (p_1',p_3'/\sqrt{\gamma})$
 obeys the linear transformation,
$$
   \pmatrix p_1'\\ p_2 \\ p_3'\endpmatrix = 
  \Gamma_+
   \pmatrix p_1\\ p_2 \\ p_3\endpmatrix , \qquad
  \Gamma_+:=\pmatrix \beta &0 &\sqrt{1-\beta^2}  \\
     0 &  1  &0\\
      \sqrt{1-\beta^2} & 0&-\beta  \\
     \endpmatrix .\tag 16
$$
Similarly the collision  
$(p_2,p_3/\sqrt{\gamma}) \to (p_2',p_3'/\sqrt{\gamma})$
 is given by
$$
   \pmatrix p_1\\ p_2' \\ p_3'\endpmatrix = 
   \Gamma_-
    \pmatrix p_1\\ p_2 \\ p_3\endpmatrix ,\qquad
   \Gamma_-:=\pmatrix 
  1 & 0 & 0\\ 
    0 & \beta &\sqrt{1-\beta^2}  \\
    0&   \sqrt{1-\beta^2} & -\beta
    \endpmatrix. \tag 17
$$
From ref.[1,2], the transition from the initial sate 
$(k_1,0,0)$ to the final state $(0,k_1,0)$
occurs which is expressed as
$$
   \pmatrix0\\k_1 \\ 0\endpmatrix 
   =\Gamma_- \Gamma_+ \Gamma_- \Gamma_+
    \cdots \Gamma_- \Gamma_+ \Gamma_- \Gamma_+
    \pmatrix k_1 \\ 0 \\ 0 \endpmatrix=
     \left(\Gamma_- \Gamma_+\right)^{n}
      \pmatrix k_1 \\ 0 \\ 0 \endpmatrix ,
               \tag 18
$$
if the mass ratio is given by
$$
   \gamma=-\frac{\cos \left(\frac{2\pi n}{2n+1}\right)}
  {\cos \left(\frac{2\pi n}{2n+1}\right)+1},
            \tag 19
$$
where $n$ is a fixed integer parameter.
It should be remarked  that (19) contains $\gamma=1$
 case as $n=1$.
Thus in the classical particle chain system, (19) is
 a generalization of $\gamma=1$ case
in which the initial momentum $k_1$ of the first particle
 are completely transferred to 
that of the second particle and such transition propagates
 like a soliton [1,2].
In fact, recently a solitary wave was found for
 an  initial condition in
 classical chain system with general $n$ mass ration. 
 
Due to the time reversion symmetry of the newtonian mechanics,
 the relation also holds for (19),
$$
    \pmatrix k_2 \\ 0\\0\endpmatrix 
    = \left(\Gamma_+ \Gamma_-\right)^{n}
    \pmatrix 0 \\ k_2 \\  0 \endpmatrix.
               \tag 20
$$
One of purposes of this letter is to investigate 
a quantum system which inherits these
novel properties of the classical system.

Noting above properties, I will consider the
  three-body wave function in this quantum system
which is given by the diagram illustrated in Fig.3.
Here it should be also noted that while in the classical
 particle chain system in ref.[1,2] 
one should pay attentions upon the configuration space, 
in this quantum system (1) and (2), due to the quantum
 principle, I need not take care of the configuration
  space as long as the momentum representation.

Since $\phi$ particle does not interact with itself,
 only the loop diagram appears consisting 
of only $\phi$ and the first $\Phi$ particles or $\phi$
 and the second $\Phi$
particles. Since in the loop diagram,
 one does not need impose the 
energy-momentum conservation, the loop diagram does not
 generates  new momentum ratio as I showed in  (12). 
On the other hand, even in the virtual states,
 the momentum and the energy must be conserved 
in  the intermediate states except the loop diagrams.
 This situation is essentially the same as that of the 
 calculus of the Bethe ansatz for the case of $\gamma=1$.
In the "collision" graph of Fig.3, at which the thin line
 and thick line are incoming
and  outgoing, the momentum transfer occurs. 
For the  initial condition  $(k_1,k_2,k_3/\sqrt{\gamma})$,
 in the perturbative calculation, 
there appear infinite types of momentum
$$
 (\Gamma_-\Gamma_+)^m\pmatrix k_1 \\ k_2 \\ k_3 \endpmatrix,
  \quad
 \Gamma_+(\Gamma_-\Gamma_+)^m
 \pmatrix k_1 \\ k_2 \\ k_3 \endpmatrix,
               \tag 21
$$
where $m$ is a non-negative integer. 
Thus in general this system cannot be exactly calculated
 by the perturbation scheme
and thus neither appear integrable.

However when the mass ratio is given by (19), 
the kinds of momentums appearing in the perturbative
 calculation becomes, at most, $4n+1$. 
In other words, I obtain the relation
$$
     (\Gamma_-\Gamma_+)^{2n+1} = 1. \tag 22
$$
The relation (22) is proved as follows;
$$
   \split
   &(\Gamma_-\Gamma_+)^{2n+1}
     \pmatrix k_1 \\ k_2 \\ k_3 \endpmatrix\\
   &=(\Gamma_-\Gamma_+)^{n+1}(\Gamma_-\Gamma_+)^{n}
    \pmatrix k_1 \\ k_2 \\ k_3\endpmatrix
  =\Gamma_-(\Gamma_+\Gamma_-)^{n}\Gamma_+
  \pmatrix \cdot \\ k_1 \\ \cdot\endpmatrix
    =\pmatrix  k_1 \\ \cdot \\ \cdot\endpmatrix\\
    &=(\Gamma_-\Gamma_+)^{n}\Gamma_-(\Gamma_+\Gamma_-)^{n}
    \Gamma_+\pmatrix k_1 \\ k_2 \\ k_3\endpmatrix
  =(\Gamma_-\Gamma_+)^{n}\Gamma_-\pmatrix k_2 \\ \cdot
    \\ \cdot\endpmatrix
   =\pmatrix  \cdot \\k_2 \\ \cdot\endpmatrix\\
    &=\pmatrix k_1 \\ k_2 \\ \cdot \endpmatrix
      .\endsplit \tag 23
$$
where  $\cdot$'s are unknown quantities. 
Here I used  (18), (20) and the relations;
$$
         \Gamma_+\pmatrix \cdot\\ k_2\\ \cdot\endpmatrix
          = \pmatrix \cdot\\ k_2\\ \cdot\endpmatrix,\quad
         \Gamma_-\pmatrix k_1\\ \cdot\\  
         \cdot\endpmatrix=\pmatrix k_1\\ \cdot\\ 
          \cdot\endpmatrix,
         \quad \Gamma_+^2=\Gamma_-^2=1. \tag 24
$$
By the energy-momentum conservation, the third component 
in (23) is determined as $k_3$
and $(\Gamma_-\Gamma_+)^{2n+1}$ must be unit matrix.

As the transitions between the momentums of the first and
 second $\Phi$ particles occur $n$ times in the perturbative
  diagram mediated by the $\phi$ particle, 
the momentum completely exchanges;
 $(k_1,\cdot,\cdot) \to (\cdot,k_1,\cdot)$
and similarly $(\cdot,k_2,\cdot)\to (k_2,\cdot,\cdot)$.
After $(4n+2)$ "collisions" in the calculus, the initial
 momentum $(k_1,k_2,k_3)$ recovers.
Since such shuttles of the momentum repeat in the calculus
 of the perturbation scheme,
 there appear, at most, $4n+1$ kinds of momentums 
 in the calculation.
Accordingly if the mass ratio given by (19),
 three-body wave function is classified by, at most,  $4n+1$ 
momentum states and can be  calculated in principal
 like that in ref.[6].
Due to the relation (12), it is  expected that the phase
 factor consists of the (pseudo-)velocities.

\subheading{Discussion}

In (13), there appears the ratio $k_1/(\sqrt{\gamma}k_2)$
 in the phase of collision.
This ratio corresponds to the ratio of the velocity rather
 than the momentum of the classical particles.
  Hence I emphasis that the quasi-momentum in [3] should be
interpreted as the modification of the velocity.
 I believe that this picture plays an important role if one
  goes beyond the integrable model to more realistic model.

Furthermore using the diagrams, I conjectured that 
the three-body wave function  of three particles can be
 exactly calculated for the special mass 
ratio which was calculated in ref.[1,2]. 
Thus  this quantum system partially succeeds the properties of
  solitary wave in 
the classical system of the hard core chain [1,2].

Further computations will be reported somewhere [7].

\subheading{Acknowledgment}

I am grateful Prof. K. Tamano and W. Kawase for helpful
 discussions for the quantum integrable system at far earlier
  stage of this study.
I thank Dr. S. Ishiwata and Prof. S.  Saito for drawing my
 attention to this problem.

\vskip 0.5 cm

\subheading{Figure Captions}

Fig.1: Sum of graphs for the two-body wave function.

Fig.2: Resonance of two-body wave function.

Fig.3: Sum of graphs for the three-body wave function.

\enddocument